# Alternative understanding of the skyrmion Hall effect based on one-dimensional domain wall motion

*Kyoung-Woong Moon,[1] Jungbum Yoon,[1] Changsoo Kim,[1] Jae-Hun Sim,[2] Se Kwon Kim[3] Soong-Geun Je,[2,*] and Chanyong Hwang[1,*]*

[1]Quantum Spin Team, Korea Research Institute of Standards and Science, Daejeon 34113, Republic of Korea

[2]Department of Physics, Chonnam National University, Gwangju 61186, Republic of Korea

[3]Department of Physics, Korea Advanced Institute of Science and Technology, Daejeon 34141, Republic of Korea

*Correspondence: sg.je@jnu.ac.kr and cyhwang@kriss.re.kr

**Abstract**

A moving magnetic skyrmion exhibits transverse deflection. This so-called skyrmion Hall effect has been explained by the Thiele equation. Here, we provide an alternative interpretation of the skyrmion Hall effect based on the dynamics of domain walls enclosing the skyrmion. We relate the spin-torque-induced local rotation of the domain wall segments to the shift of the skyrmion core, explaining the skyrmion Hall effect at the micromagnetic level. Bases on our intuitive interpretation, we also show that the skyrmion Hall effect can be suppressed by combining the spin-transfer and spin-orbit torques, whereby removing the major obstacle to utilizing skyrmions in devices.

A magnetic skyrmion (SK) is a whirling spin texture of which topology is defined by the unit skyrmion number that serves as the unit topological charge. The presence of the finite topological charge gives rise to novel features that originate from emergent electromagnetism such as the topological Hall effect or the skyrmion Hall effect (SKHE)[1-5]. As the nomenclature reflects, the SKHE refers to the transverse motion of the skyrmion structure when it is forced to move in the logitudinal direction, analogous to the Magnus effect in classical mechanics[6-8] that deflects the motion of a rotating object from the direction of the force.

The SKHE is intuitively understood as the counter-effect of the topological Hall effect[1,6]. When conduction electrons pass through a SK structure, they experience the emergent magnetic field. The emenrgent magnetic field in turn leads to the deflection of conduction electrons by the Lorentz force, bringing about the topological Hall effect. By the law of action-reaction, the SK is then deflected in the oppodite difection of the topological Hall effect of conduction electrons. Meanwhile, the SK motion driven by spin torques has been qualitatively described by Thiele equation that is obtained by the certain form of the spatial average of the Landau-Lifshits-Gilbert (LLG) equation for a rigid skyrmion structure which only displays translational motion[7,9-11].

The Thiele equation for the rigid Néel-type SK motion driven by the damping-like spin-orbit torque reads[7,11]

$$\mathbf{G} \times \mathbf{v} - \alpha \overleftrightarrow{\mathbf{D}} \cdot \mathbf{v} + B\mathbf{j}_{HM} = 0, \qquad (1)$$

where $\mathbf{G} = (0,0,-4\pi Q)$ is the gyro-coupling vector with the topological charge $Q = \pm 1$, $\mathbf{v}$ is the drift velocity of the SK, $\alpha$ is the Gilbert damping constant, $\overleftrightarrow{\mathbf{D}} = \begin{pmatrix} D & 0 \\ 0 & D \end{pmatrix}$ is the dissipative force tensor, $B$ is the coefficient which depics the direction and the strength of the driving force due to the damping-like torque, and $\mathbf{j}_{HM} = (j_{HM}, 0)$ is the electric current density generating the damping-like spin-orbit torque. The gyro-coupling term ($\propto G$) is responsible for the SKHE, and Eq. (1) leads to the SK velocities

$$V_x = \frac{\alpha D}{G^2 + \alpha^2 D^2} Bj_{HM} \qquad (2a)$$

$$V_y = \frac{G}{G^2 + \alpha^2 D^2} Bj_{HM}. \qquad (2b)$$

While the Thiele equation predicts the SKHE by the presence of the transverse velocity $V_y$, the detailed mechanism of how the $V_y$ arises at the micromagnetic level is not explicitlly depicted in the Thiele equation. Specifically, the SK motion is caused by the spin torques acting on the domain wall (DW) enclosing the SK, but the relationship between the SKHE and DW dynamics is lacking in the Thiele equation. Technologically, understanding the SKHE based on the DW dynamics is important because the current-induced DW motion which have been actively studied for the DW racetrack memory[12-14] can directly be applied to the SK-based devices[9,15], for example, in suppressing the SKHE in a device[16-23].

Here we suggest an alternative explanation of the SKHE based on the DW dynamics. In the one-dimensinal (1-D) model of the DW, the spin-torques result in a shift of the DW position as well as rotation of the DW magnetization[24,25]. We apply the one-dimensional model to the DW segements of the SK and relate the local nonuniform rotations of the DW segements to the shift of the SK core that can be interpreted as the SKHE. Our approach also suggests a novel condition for eliminating the SKHE by combining the spin-transfer torque (STT) and the spin-orbit torque (SOT) that is confirmed by micromagnetic simulation. These results allow a more intuitive understanding of the SK dynamics and provide useful information for the development of SK-based devices.

Fig. 1a depicts the system which hosts the DWs and SKs. We assume that the electron flow is in the *x* direction, and then the electron flows in the ferromagnetic (FM) layer and the heavy metal (HM) layer induce the STT and SOT, respectively. The dynamics of magnetization in this system is governed by the Landau–Lifshitz–Gilbert (LLG) equation as follows[13,24,25]:

$$\dot{\mathbf{m}} = -\gamma \mathbf{m} \times \mathbf{H}_{\text{eff}} + \alpha \mathbf{m} \times \dot{\mathbf{m}} - u\partial_x \mathbf{m} + \beta u \mathbf{m} \times \partial_x \mathbf{m} - \gamma \tau_d \mathbf{m} \times (\mathbf{m} \times \boldsymbol{\sigma}). \qquad (3)$$

The torque terms on the right side of the equation are, from the left, a field torque, a damping torque, an adiabatic STT, a nonadiabatic STT[26-28], a damping-like SOT[13,28-31]. Here, $\mathbf{m}$ is the normalized local magnetization vector, $\gamma$ is the gyromagnetic ratio, $\mathbf{H}_{\text{eff}}$ is the effective field, $\alpha$ is the damping constant, $u$ is the strength of STT, i.e., the spin-polarized current density, $\beta$ is the nonadiabaticity of STT, $\tau_d$ is the strength of the damping-like SOT, and $\boldsymbol{\sigma}$ is the direction of the pumped spin moment, generated by the spin Hall effect[29,30]. The $\boldsymbol{\sigma}$ is responsible for the damping-like SOT. In Eq. (3), we neglected the field-like SOT[31] because the field-like torque is negligibly small in the present system[30]. In our system, positive values of $u\beta$ and $\tau_d$ drive DWs and SKs to the +$x$ direction.

The motions of the DW and SK (illustrated in Fig. 1b) can be obtained by using mumax3 micromagnetic simulation code[32] which numerically solves the LLG equation. Fig. 1c shows the positions ($q_x$ and $q_y$ in Fig. 1b) of the DW and the SK as a function of time after turning on the current $u = 100$ m/s. To convey the essential idea of how the DW dynamics can be used to explain the SKHE, we set the external field, $\beta$, and $\tau_d$ to be zero. This means that only the adiabatic STT is considered. For more information on the simulation parameters, please see the caption of Fig. 1c. While the position of the DW slightly changes at first and finally converges to a certain value (the DW stops), the positions of the SK constantly increase with time with a constant speed. These distinct behaviors of the DW and the SK under the same condition can be explained by the 1-D DW equation as follows.

The 1-D model for the DW motion can be obtained by inserting the DW profile $\mathbf{m}(x)$ into Eq. (3). The simplest description of the DW is $\mathbf{m}(x) = (\sin\theta\cos\psi, \sin\theta\sin\psi, \cos\theta)$, where $\cos\theta(x) = -1 \times \tanh[(x - q_x)/\Delta_0]$, $\psi$ is the DW magnetization angle, $q_x$ is the DW position, and $\Delta_0$ is the DW width (Fig. 1b). The 1-D model reads[13,24,25]

$$\dot{q}_x = \frac{\Delta_0}{\alpha}\left[\gamma H_z + \frac{\beta u}{\Delta_0} + \frac{\gamma\pi\tau_d \cos\psi}{2} - \dot{\psi}\right], \quad (4a)$$

$$(1 + \alpha^2)\dot{\psi} = \gamma H_z + \frac{u(\beta - \alpha)}{\Delta_0} + \frac{\gamma\pi\tau_d \cos\psi}{2} - \frac{\gamma\alpha\pi H_{\mathrm{DMI}} \sin\psi}{2}. \quad (4b)$$

Here, $H_z$ is the perpendicular field, $H_{\mathrm{DMI}}$ is the effective field from the Dzyaloshinskii-Moriya interaction (DMI), acting on the DW[25,33]. Strong positive $H_{\mathrm{DMI}}$ leads to the chiral Néel DW and SK structures shown in Fig. 1b.

Fig. 1d schematically clarifies the reason for the eventual stopping of the DW. When $H_z = \beta = \tau_d = 0$, Eq. (4a) tells that the DW position $q_x$ only depends on the variation of $\psi$. In Eq. (4b), the adiabatic STT can cause the rotation of $\psi$, $-u\alpha/\{\Delta_0(1 + \alpha^2)\}$ ($= \dot{\psi}_{\mathrm{adia}}$), and the $\dot{\psi}_{\mathrm{adia}}$ induces nonzero $\dot{q}_x$ as well as the rotation of the DW magnetization from the static Néel DW structure (change in $\psi$) at the initial state. However, the change in $\psi$ by $\dot{\psi}_{\mathrm{adia}}$ in turn produces a nonzero $\gamma\alpha\pi H_{\mathrm{DMI}}\sin\psi/\{2(1 + \alpha^2)\}$ ($= \dot{\psi}_{\mathrm{DMI}}$), and, when the DW angle $\psi$ reaches a certain value, the $\dot{\psi}_{\mathrm{DMI}}$ cancels out the $\dot{\psi}_{\mathrm{adia}}$, resulting the net change of the DW angle vanishes $\dot{\psi} = \dot{\psi}_{\mathrm{adia}} + \dot{\psi}_{\mathrm{DMI}} = 0$. Then the coupling between $\dot{\psi}$ and $\dot{q}_x$ in Eq. (4a) results in $\dot{q}_x = 0$. This explains why the DW finally stops with the slight rotation of the DW magnetization after the initial displacement in Fig. 1c.

In contrast to the DW case, however, the SK can continuously move without stopping. Fig. 1e describes the SK case. The magnetization configuration of the SK with pink arrows in the red dashed box resembles the DWs shown in Figure 1d. Therefore, one can expect that the same dynamics occurs at the beginning of the SK motion; shift of the DW ($q_x$) is induced by the rotation of $\psi$ (gray arrows in the red dashed box). It is noteworthy that the gray arrows in the red dashed box are indeed a part of the SK with gray arrows. In other words, different from the DW case, the local change in $\psi$ can be compensated for by the shift of the SK center to the $y$ direction. Even after the initial shift of the SK, there is still a region where the $\dot{\psi}_{\text{adia}}$ is not compensated by the $\dot{\psi}_{\text{DMI}}$ (gray arrows in the white dashed box), leading to a successive shift of the SK. From the above descriptions, one can conclude that the transverse motion of the SK originates from the local rotation of the DW segments enclosing the SK, and this explanation constitutes the intuitive explanation of the SKHE from the micromagnetic point of view.

To calculate the skyrmion velocity using the DW dynamics, one has to consider the main difference between the DW and the SK; the SK contains all the DW tilting angles with respect to the electric current direction. Fig. 2a shows the magnified view of a DW segment that forms the SK with the definition of coordinates. Note that $\varphi$ is different from $\psi$. In the previous 1-D DW equation, $\varphi$ is just 0 or $\pi$. Now, we use the generalized DW equations that include the DW tilting angle $\varphi$ as a new variable[28,34]. The resultant equations are given by

$$\dot{q}_x = \frac{\Delta_0}{\alpha \cos \varphi} \left[ \gamma H_z + \frac{\beta u \cos}{\Delta_0} + \frac{\gamma \pi \tau_d \cos \psi}{2} - \dot{\psi} \right], \tag{5a}$$

$$(1 + \alpha^2)\dot{\psi} = \gamma H_z + \frac{u(\beta - \alpha) \cos \varphi}{\Delta_0} + \frac{\gamma \pi \tau_d \cos}{2} - \frac{\gamma \alpha \pi H_{\text{DMI}} \sin(\psi - \varphi)}{2}. \tag{5b}$$

Eq. (5a) contains two $\cos \varphi$ terms. The first $\cos \varphi$ before the bracket represents the coordinate rotation. Since the $H_z$ pushes the DW to its normal direction, converting the $\dot{q}_x$ in Eq. (4a) to the speed along the $x$-direction requires the $\cos \varphi$ (Fig. 2a). The second $\cos \varphi$ inside the bracket is introduced to take into account the DW normal component of the electron flow ($u$) in the $x$-direction[35].

Now we will roughly derive the equation for the SK motion using Eq. (5). As schematically shown in Fig. 2b, the infinitesimal change of the SK position in the $y$ direction ($\delta q_y$) is related to infinitesimal change of the DW magnetization angle ($\delta\psi$) at $\varphi$ such that $\delta q_y = -r_0 \cos\varphi\, \delta\psi$ where $r_0$ is the radius of the SK. Because the SK consists of DW segments along the SK boundary, integration of Eq. (5) from $\varphi=0$ to $\varphi=2\pi$, with an assumption that the DMI is strong so that $\psi \approx \varphi$, results in the SK velocity equations as follows.

$$V_x \approx \frac{\alpha}{1+\alpha^2}\left[\frac{u}{\alpha} + \beta u + \frac{\gamma\pi\Delta_0\tau_d}{2}\right]. \tag{6a}$$

$$V_y \approx -\frac{r_0}{2(1+\alpha^2)}\left[\frac{(\beta-\alpha)u}{\Delta_0} + \frac{\gamma\pi\,\tau_d}{2}\right]. \tag{6b}$$

When calculating $V_y$, $\delta q_y = -r_0 \cos\varphi\, \delta\psi$ is particularly considered. However, when calculating $V_x$, we neglected such additional conversion $\delta q_x = r_0 \sin\varphi\, \delta\psi$ because, around $\varphi = \pi/2$ or $3\pi/2$, where the conversion is maximized, the spin torques vanish, giving rise to a negligible contribution. These equations have similar forms to the SK equations derived the from the Thiele equation[10,11,19] and also predict the SKHE well. Fig. 2c presents the SK velocities as a function of $\beta$ when $\tau_d = 0$. A good agreement between the results from Eq. (6) (solid lines) and those of micromagnetic simulations (symbols) confirms the validity of our approach.

However, we would like to note that our approach is simplified treatment so that the agreement between the damping-like SOT terms in Eq. (6) and Eq. (2) is not perfect. For example, according to Eq. (2), $V_y/V_x \propto 1/D \propto \Delta_0/r_0$[7,11], but $V_y/V_x \propto r_0/\Delta_0$ in Eq. (6) for the SOT contribution. We ascribe the disagreement to our oversimplified treatment of the SOT. Different from the STT, which only depends on $\varphi$, the SOT term in Eq. (5b) depends on $\psi$, affecting $\dot{\psi}$ and vice versa. This complex aspect of the SOT might not be handled in our simple approach thoroughly, requiring further modification. Nevertheless, we emphasize that our approach correctly predicts the direction of the SKHE for each spin torque entirely in terms of the dynamics of a DW without invoking the Thiele equation and provides intuitive understanding of the SKHE from the micromagnetic point of view.

Finally, we discuss an alternative way of suppressing the SKHE. The SKHE has been recognized as one of the challenges in developing the SK-based device since it constantly deflects the skyrmion motion from the desired direction. Therefore, numerous studies have been undertaken to develop ways to suppress the transverse motion; approaches include using a confined track[16-18], an antiferromagnetic SK[19,20], a skyrmionium[21,22], and a hybrid-DMI[23]. According to our approach, we find that the SKHE originates from the local magnetization rotation in the DW parts. In other words, one can expect the elimination of the SKHE when $\dot{\psi} = 0$ is satisfied. If we assume that $H_z = \tau_d = 0$, according to Eq. (5b) and Eq. (6b), the condition is simply given by $\alpha = \beta$. This situation is already shown in Fig. 2c where the $V_y$ becomes zero when $\alpha = \beta$.

However, most of the multilayers[11,29,34] are expected to have negligibly small $\beta$ except for special cases[27]. In addition, it is hard to achieve the $\alpha = \beta$ condition by tuning the multilayer structures. Thus, considering both the SOT and STT can provide a more feasible way because $\tau_d$ is tunable by engineering material parameters[13,30]. For instance, $\tau_d$ contains the spin Hall effect efficiency, which differs for different adjacent heavy metals, the saturation magnetization, and the magnetic layer thickness. Fig. 3a summarizes the simulation results that show the SK trajectories for different $\tau_d$ under the same $u$ and $\beta$ ($\neq \alpha$). It is interesting that, for the given STTs, the SKHE direction can change depending on the strength of SOT, and a certain value of $\tau_d$ exactly removes $V_y$. When the $H_{\text{DMI}}$ is strong enough to fix the DW magnetization, the DW magnetization direction is normal to the local DW; that is, $\psi \approx \varphi$. Inserting $\psi \approx \varphi$ in Eq. (5b), one can obtain a condition $\frac{(\alpha-\beta)}{\Delta_0}u = \frac{\gamma\pi}{2}\tau_d$ leads to $\dot{\psi} = 0$ when $H_z = 0$. This condition also yields $V_y = 0$ in Eq. (6b). What happens at this condition is that the contributions from the damping-like SOT, the adiabatic STT, and the nonadiabatic STT to $\dot{\psi}$ are completely compensated, resulting in the suppression of the SKHE. We emphasize that this method is completely different from the previously proposed schemes[16-23]. Fig. 3b shows the required $\tau_d$ for eliminating $V_y$ as a function of $u$ for several $\beta$ values. The solid lines are obtained by the condition $\frac{(\alpha-\beta)}{\Delta_0}u = \frac{\gamma\pi}{2}\tau_d$, and they are consistent with micromagnetic simulation data (symbols). The linear relation between the required $\tau_d$ and $u$ is important because $\tau_d$ is proportional to the total current density in a given structure. This means that the condition for $V_y = 0$ does not rely on the amount of the total current.

To summarize, we studied the micromagnetic origin of transverse motion of the SK by using a simple 1-D DW model without invoking the Thiele equation. We relate the DW magnetization rotation by spin torques to the SK center shift in the transverse direction, explaining the SKHE in terms of micromagnetics. Also, we find a condition for which the angle of DW magnetization does not change from the initial value during the motion. When such a condition is satisfied, the SK motion follows the simple DW motions with no transverse motions of the SK. Our findings provide intuitive understanding of the SKHE and also useful conditions for the development of SK-based devices.

**References**

1. N. Nagaosa, and Y. Tokura, Nat. Nanotechnol. **8**, 899 (2013).

2. A. Fert, N. Reyren, and V. Cros, Nat. Rev. Mater. **2**, 17031 (2017).

3. W. Jiang, G. Chen, K. Liu, J. Zang, S. G. E. T. Velthuis, and A. Hoffmann, Phys. Rep. **704**, 1 (2017).

4. G. Finocchio, F. Büttner, R. Tomasello, M. Carpentieri, and M. Kläui, J. Phys. D. Appl. Phys. **49**, 423001 (2016).

5. S. Woo, K. Litzius, B. Krüger, M.-Y. Im, L. Caretta, K. Richter, M. Mann, A. Krone, R. M. Reeve, M. Weigand, P. Agrawal, I. Lemesh, M.-A. Mawass, P. Fischer, M. Kläui, and G. S. D. Beach, Nat. Mater. **15**, 501 (2016).

6. K. Everschor-Sitte, and M. Sitte, J. Appl. Phys. **115**, 172602 (2014).

7. W. Jiang, X. Zhang, G. Yu, W. Zhang, M. B. Jungfleisch, J. E. Pearson, O. Heinonen, K. L. Wang, Y. Zhou, A. Hoffmann, and S. G. E. T. Velthuis, Nat. Phys. **13**, 162 (2017).

8. K. Litzius, I. Lemesh, B. Krüger, P. Bassirian, L. Caretta, K. Richter, F. Büttner, K. Sato, O. A. Tretiakov, J. Förster, R. M. Reeve, M. Weigand, I. Bykova, H. Stoll, G. Schütz, G. S. D. Beach, and M. Kläui, Nat. Phys. **13**, 170 (2017).

9. A. A. Thiele, Phys. Rev. Lett. **30**, 230 (1973).

10. J. Iwasaki, M. Mochizuki, and N. Nagaosa, Nat. Commun. **4**, 1463 (2013).

11. R. Tomasello, E. Martinez, R. Zivieri, L. Torres, M. Carpentieri, and G. Finocchio, Sci. Rep. **4**, 6784 (2014).

12. S. S. P. Parkin, M. Hayashi, and L. Thomas, Science **320**, 190 (2008).

13. S. Emori, U. Bauer, S.-M. Ahn, E. Martinez, and G. S. D. Beach, Nat. Mater. **12**, 611 (2013).

14. S.-H. Yang, K.-S. Ryu, and S. Parkin, Nat. Nanotechnol. **10**, 221 (2015).

15. J. Sampaio, V. Cros, S. Rohart, A. Thiaville, and A. Fert, Nat. Nanotechnol. **8**, 839 (2013).

16. I. Purnama, W. L. Gan, D. W. Wong, and W. S. Lew, Sci. Rep. **5**, 10620 (2015).

17. J. Iwasaki, M. Mochizuki, and N. Nagaosa, Nat. Nanotechnol. **8**, 742 (2013).

18. J. Iwasaki, W. Koshibae, and N. Nagaosa, Nano Lett. **14**, 4432 (2014).

19. X. Zhang, Y. Zhou, and M. Ezawa, Nat. Commun. **7**, 10293 (2016).

20. H. Xia, C. Jin, C. Song, J. Wang, J. Wang, and Q. Liu, J. Phys. D. Appl. Phys. **50**, 505005 (2017).

21. X. Zhang, J. Xia, Y. Zhou, D. Wang, X. Liu, W. Zhao, and M. Ezawa, Phys. Rev. B **94**, 094420 (2016).

22. M. Finazzi, M. Savoini, A. R. Khorsand, A. Tsukamoto, A. Itoh, L. Duò, A. Kirilyuk, T. Rasing, and M. Ezawa, Phys. Rev. Lett. **110**, 177205 (2013).

23. K.-W. Kim, K.-W. Moon, N. Kerber, J. Nothhelfer, and K. Everschor-Sitte, Phys. Rev. B **97**, 224427 (2018).

24. A. Thiaville, Y. Nakatani, J. Miltat, and Y. Suzuki, Europhys. Lett. **69**, 990 (2005).

25. A. Thiaville1, S. Rohart, É. Jué, V. Cros, and A. Fert, Europhys. Lett. **100**, 57002 (2012).

26. K.-J. Kim, J.-C. Lee, S.-J. Yun, G.-H. Gim, K.-S. Lee, S.-B. Choe, and K.-H. Shin, Appl. Phys. Express **3**, 083001 (2010).

27. S.-G. Je, S.-C. Yoo, J.-S. Kim, Y.-K. Park, M.-H. Park, J. Moon, B.-C. Min, and S.-B. Choe, Phys. Rev. Lett. **118**, 167205 (2017).

28. K.-W. Moon, C. Kim, J. Yoon, J. W. Choi, D.-O. Kim, K. M. Song, D. Kim, B. S. Chun, and C. Hwang, Nat. Commun. **9**, 3788 (2018).

29. K.-S. Ryu, L. Thomas, S.-H. Yang, and Stuart Parkin, Nat. Nanotechnol. **8**, 527 (2013).

30. P. P. J. Haazen, E. Murè, J. H. Franken, R. Lavrijsen, H. J. M. Swagten, and B. Koopmans, Nat. Mater. **12**, 299 (2013).

31. K. Garello, I. M. Miron, C. O. Avci, F. Freimuth, Y. Mokrousov, S. Blügel, S. Auffret, O. Boulle, G. Gaudin, and P. Gambardella, Nat. Nanotechnol. **8**, 587 (2013).

32. A. Vansteenkiste, J. Leliaert, M. Dvornik, M. Helsen, F. Garcia-Sanchez, and B. V. Waeyenberge, AIP Adv. **4**, 107133 (2014).

33. S.-G. Je, D.-H. Kim, S.-C. Yoo, B.-C. Min, K.-J. Lee, and S.-B. Choe, Phys. Rev. B **88**, 214401 (2013).

34. O. Boulle, S. Rohart, L. D. Buda-Prejbeanu, E. Jué, I. M. Miron, S. Pizzini, J. Vogel, G. Gaudin, and A. Thiaville, Phys. Rev. Lett. **111**, 217203 (2013).

35. K.-W. Moon, D.-H. Kim, S.-C. Yoo, C.-G. Cho, S. Hwang, B. Kahng, B.-C. Min, K.-H. Shin, and S.-B. Choe, Phys. Rev. Lett. **110**, 107203 (2013).

**Acknowledgments**

This work was supported by the Future Materials Discovery Program through the National Research Foundation of Korea (NRF-2021M3F3A2A01037663, NRF-2022M3H4A1A04071154) from the Korean government (MSIP). J.H.S. and S.G.J. were supported by the National Research Foundation of Korea grant funded by the Korea government (NRF-2020R1C1C1006194). S.K.K. was supported by the Brain Pool Plus Program through the National Research Foundation of Korea funded by the Ministry of Science and ICT (NRF-2020H1D3A2A03099291) and National Research Foundation of Korea funded by the Korea Government via the SRC Center for Quantum Coherence in Condensed Matter (NRF-2016R1A5A1008184).

**Figure captions**

**FIG. 1. (a)** Electric current flow in the FM/HM bilayer. **(b)** The DW and SK structures. Gray color represents normalized perpendicular component of magnetization ($m_z$). Pink arrows show local magnetization directions. **(c)** Simulation results of positions of the DW and the SK after turning the electric current on. In the simulations the following magnetic parameters are also used: $M_S$ (saturation magnetization, $580 \times 10^3$ A/m, $A$ (exchange stiffness constant, $1.5 \times 10^{-11}$ J/m), $K$ (perpendicular anisotropy constant, $800 \times 10^3$ J/m$^3$), $D$ (strength of the Dzyaloshinskii-Moriya interaction, $3.5 \times 10^{-3}$ J/m$^2$), $\alpha$ (damping constant 0.3). The simulation area for the DW motion is 600 nm × 20 nm × 0.4 nm. For the SK motion, the simulation area is 600 nm × 600 nm × 0.4 nm. For both DW and SK simulations, the cell size 2 nm × 2 nm × 0.4 nm is taken, and we used the periodic boundary condition in the $x$ and $y$ directions. **(d)** Illustration of the DW motions induced by the adiabatic STT. Green arrows depict the direction of $H_{\text{DMI}}$. The yellow and red arrows represent rotation of the DW magnetization angle ($\psi$) due to $H_{\text{DMI}}$ and the adiabatic STT, respectively. **(e)** Illustration of the SK motion by the adiabatic STT. Pink arrows in the x–y plane show the initial SK. Gray arrows depict the shifted SK. $V$ is the velocity of the SK.

**FIG. 2. (a)** DW structure with the tilting angle $\varphi$ and pictorial description of parameters. **(b)** Schematic illustration of the relation between $\delta q_y$ and $\delta \psi$. **(c)** Velocity of the SK as a function of $\beta$. Symbols represents the simulation results. Solid lines show the theoretical results in Eq. (6). The inset shows the rotation directions of $\psi$ due to the nonadiabatic STT (blue arrow) and the adiabatic STT (red arrow).

**FIG. 3. (a)** Simulation results of SK motion. SKs are overlapped on a single image with a fixed time interval (0.2 ns). Dashed lines are guides for SK motions. **(b)** Required value of $\tau_d$

to eliminate of $V_y$ for several $\beta$ values. The solid lies are obtained from the theoretical results, and dots are from simulations.

**Figures**

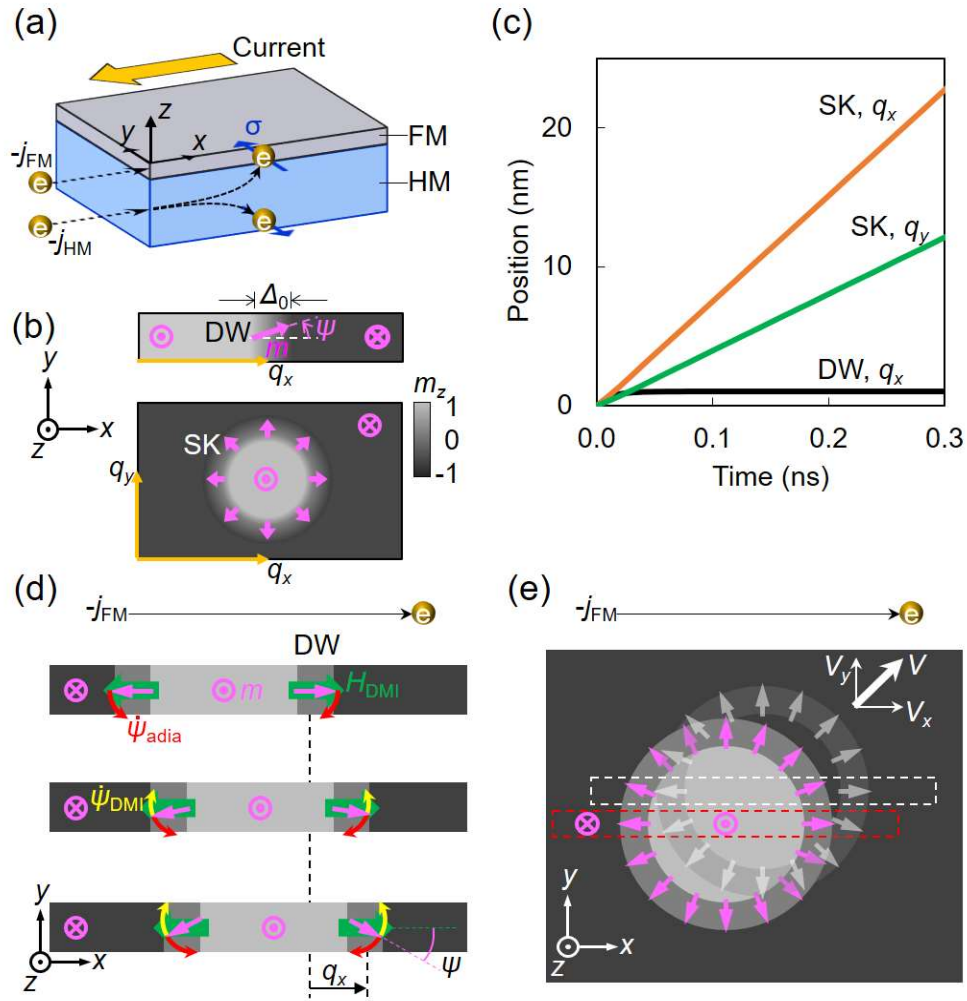

FIG. 1

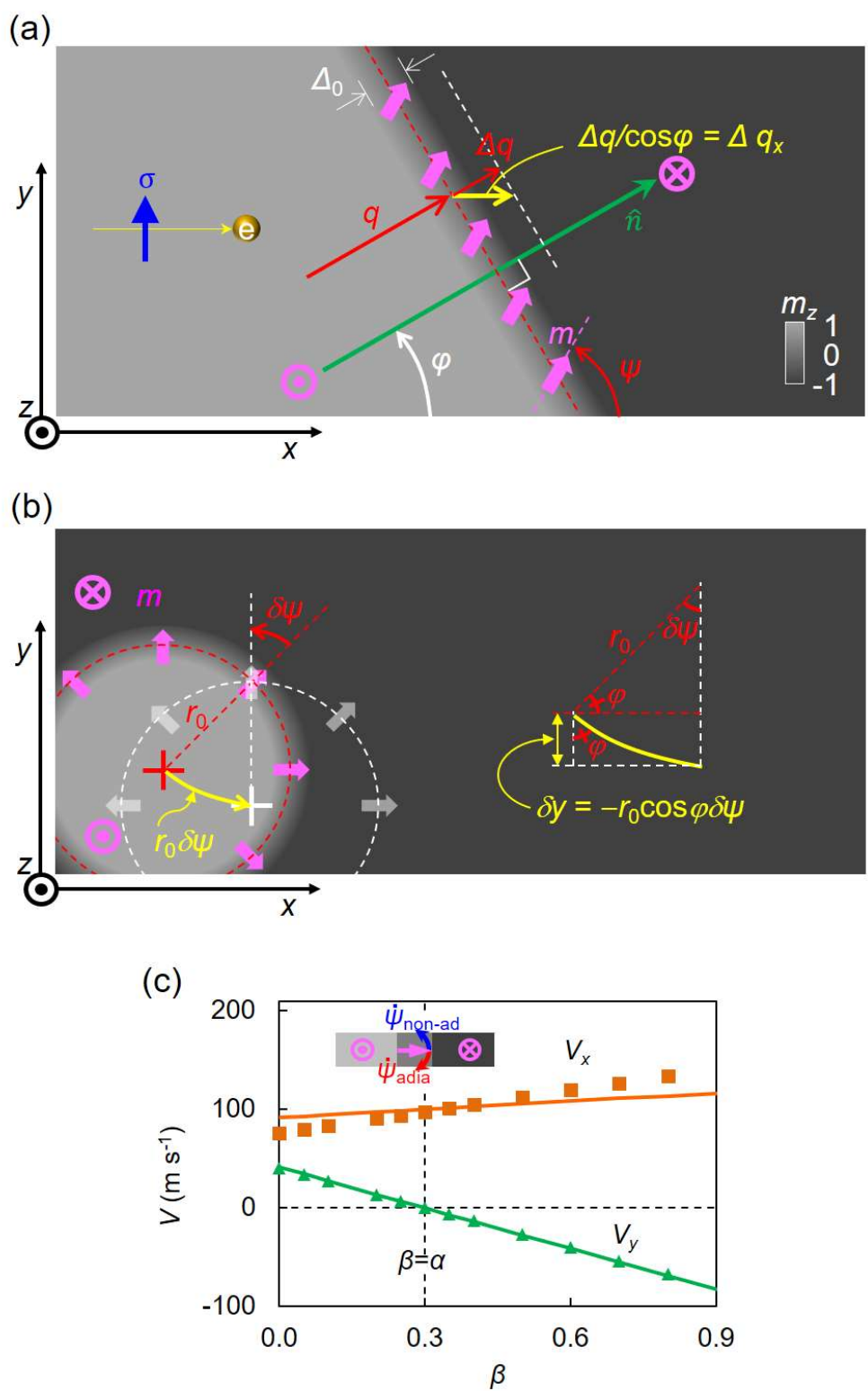

FIG. 2

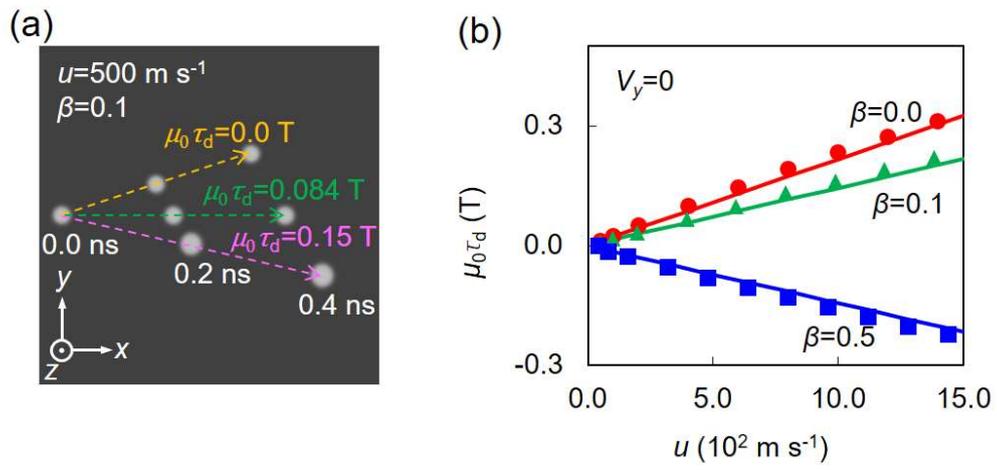

FIG. 3